\documentclass[journal]{IEEEtran}
\usepackage{cite}
\ifCLASSINFOpdf
\else
  \usepackage[dvips]{graphicx}
\fi
\usepackage{amsmath}
\usepackage{array}
\begin{document}
\title{Modulational Instability and Generation of Envelope Solitons in Four Component Space Plasmas}

\author{N. A.~Chowdhury,~A. Mannan,~M. R. Hossen,~and A. A. Mamun% <-this % stops a space
\thanks{N. A. Chowdhury, A. Mannan, and A. A. Mamun are with the Department of Physics, Jahangirnagar
University, Savar, Dhaka, Bangladesh.}% <-this % stops a space
\thanks{M. R. Hossen is with the Department of General Educational Development, Daffodil
International University, Dhanmondi, Dhaka-1207, Bangladesh.}% <-this % stops a space
\thanks{Corresponding author's e-mail:~nurealam1743phy@gmail.com}
}

\maketitle

\begin{abstract}
A four component space plasma system (consisting of immobile positive ions, inertial
cold positrons as well as hot electrons and positrons following Cairns' nonthermal distribution function is considered.
The nonlinear propagation of the positron-acoustic (PA) waves, in which the inertia (restoring force) is provided by the
cold positron species (nonthermal pressure of both hot electron and positron species) has been theoretically investigated by
deriving the nonlinear Schr\"{o}dinger (NLS) equation. It is found from the numerical analysis of this NLS equation that the
space plasma system under consideration supports the existence
of both dark and bright envelope solitons associated with PA waves, and that the dark (bright) envelope solitons are modulationally
stable (unstable). It is also observed that the basic properties (viz. stable regime and unstable regime with growth rate) of the PA envelope solitions  are
significantly modified by related plasma parameters (viz. number densities and temperature of plasma species),
which correspond to different realistic space plasma situations.
\end{abstract}

\begin{IEEEkeywords}
Positron-acoustic waves, Modulational instability, Envelope solitons.
\end{IEEEkeywords}

% \fi
\IEEEpeerreviewmaketitle

\section{Introduction}
%%%%%%%%%%%%%%%%%%%%%%%%%%%%%%%%%%%%%%%%%%%%%%%%%%%%%%%%%%%%%%%%%%%%%%%%%%%%%%%%%%%%%%
Nowadays, the physicists are mesmerized by the natural beauty of electron-positron-ion (e-p-i)
plasmas because many painstaking observations  disclosed the existence of e-p-i plasmas in
various regions of our universe (such as supernovas, pulsar environments, cluster
explosions \cite{Begelman1984,Miller1987,Tribeche2009}, etc.),  polar regions of neutron stars
\cite{Michel1991}, white dwarfs \cite{Hossen2014a,Hossen2014b}, early universe \cite{Misner1973},
inner regions of the accretion disc surrounding black holes \cite{Rees1971}, pulsar
magnetosphere \cite{Liang1998,Michel1982}, center of our galaxy \cite{Burns1983},
and solar atmospheres \cite{Goldreich1969,Hansen1988}.

To understand the physics of collective processes in such kind of
plasmas, many researchers have studied the ion-acoustic waves
(IAWs) \cite{Bains2010,Rehman2016,Lu2015} and electron-acoustic
waves (EAWs) \cite{Sultana2011,Demiray2016} in e-p-i plasmas. A
few of them  have  considered isothermal Maxwellian distribution
\cite{Haque2003,Shah2010,Sabry2009,Akbari2010,El-Awady2010}, for
their considered plasma species. But in astrophysical
environments  generally a nonthermal plasma (present of excess
non-Maxwellian particles such as electrons and positrons, which is
absent from thermodynamic equilibrium, that means at least one of
the component of such kind of plasmas  does not follow the
prominent Maxwell-Boltzmann distribution) are characterized by
long tail in the high- energy region \cite{Alam2013}, around the
Earth's bow shock \cite{Matsumoto1994}, lower part of
magetosphere\cite{Bostrom1992}, and upper Martin
ionosphere\cite{Lundin1989}. By employing nonthermal distribution
for plasma species instead Maxwellian distribution in the highly
populated nonthermal particles region, obtained result is
comparatively more acceptable with, which observed by Freja and
Viking Satellites \cite{Bostrom1992,Dovner1994}. So for better
understanding about this high energetic space plasmas nonthemal
distribution can be used to model such kind of space plasma
system.

A set of researchers have been used nonthermal distribution to
study linear and nonlinear structure of e-p-i plasmas. Like
Cairns \textit{et al.} \cite{Cairns1995} used  nonthermal
distribution of electrons to understand, how the presence of a
population of energetic electrons changes the nature of ion sound
solitary waves. By using pseudo-potential method, Pakzad
\cite{Pakzad2009} studied  that under certain criteria, the
formation of solitons and the effect of nonthermal electrons on
solitons  in e-p-i plasmas. Sahu \cite{Biswajit2010} analyzed the
effects of ion kinematics viscosity on the properties of PA Shock Waves.
Messekher \textit{et al.} \cite{Messekher2016} examined the
influence  of  quantum  effects  on solitary (quantum
positron-acoustic waves)  structures  as  well  as double-layers
by deriving Korteweg-de Vries equation in an unmagnetized  four
component plasmas. By employing the reductive perturbation method
(RPM), Eslami \textit{et al.} \cite{Eslami2011} investigated
modulational instability (MI) of IAWs in
q-nonextensive e-p-i plasmas. Sultana and Kaurakis
\cite{Sultana2011}, by making use of a multiscale perturbation
technique, a NLS equation is derived to examined the stability
of the EAWs and formation of envelope solitons under certain
conditions in e-i plasmas. Zhang \textit{et al.} \cite{Zhang2009}
studied the MI for e-p-i
plasma system and observed that the amplitude of dark and bright
envelope solitons significantly depends on the effects of
nonthermal parameter, concentration of positrons and ion
temperatures. Up to the best of our knowledge, no theoretical
investigations have been worked out about the nonlinear
properties of positron-acoustic waves (PAWs) in unmagnetized plasmas with immobile ions,
inertial cold positrons, nonthermal distributed hot electrons and
positrons. Therefore, in our present work, we attempt to study
the MI of PAWs and formation of envelope solitons by deriving
the NLS equation in a  plasma having excess of nonthermal
distributed hot electrons and positrons in a ``non-Maxwellian
tail".

The present paper is organized as follows.  The basic governing equations
of our plasma model are presented in Sec. II. By using perturbation technique,
we derive a NLS equation which governs the
slow amplitude evolution in time and space in Sec. III.  The stability
analysis is presented in Sec. IV. Envelope solitons are devoted in Sec. V. Conclusion is preserved in sec.VI.
%%%%%%%%%%%%%%%%%%%%%%%%%%%%%%%%%%%%%%%%%%%%%%%%%%%%%%%%%%%%%%%%%%%%%%%%%%%%%%%%%%%%%%%%%%%%%%%%%%%%%%%%%%%%%%
\section{governing equations}
We consider an unmagnetized four component plasma system consisting of immobile positive ions, inertial
cold positrons, nonthermally distributed hot electrons and hot positrons. At equilibrium,
the quasi-neutrality condition can be expressed as $n_{cp0}+n_{hp0}+n_{i0}=n_{e0} $,
where $ n_{cp0}, n_{hp0}, n_{i0}$, and  $n_{e0} $ are the unperturbed number
densities of cold positron, hot positron, immobile ion, and hot electron respectively.
The normalized governing equations of the PAWs in
our considered plasma system are given by
\begin{eqnarray}
&&\hspace*{-3.0cm}\frac{\partial n_{cp}}{\partial t}+\frac{\partial}{\partial x}(n_{cp} u_{cp})=0,\label{a1}\\
&&\hspace*{-3.0cm}\frac{\partial u_{cp}}{\partial t} + u_{cp}\frac{\partial u_{cp} }{\partial x}=-\frac{\partial \phi}{\partial x},\label{a2}\\
&&\hspace*{-3.0cm}\frac{\partial^2 \phi}{\partial x^2}=-n_{cp} -\mu_1 n_{hp}+\mu_2 n_e-\mu_3. \label{a3}
\end{eqnarray}
For inertialess hot positrons and hot electrons are given by the following expression,
\begin{eqnarray}
&&\hspace*{-2.2cm}n_{hp}=(1+\beta\sigma_1 \phi+\beta\sigma^2_1 \phi^2)~\mbox{exp}(-\sigma_1 \phi),\nonumber\\
&&\hspace*{-2.2cm}n_e=(1-\beta\sigma_2 \phi+\beta\sigma^2_2 \phi^2)~\mbox{exp}(\sigma_2 \phi). \label{Ch1}
\end{eqnarray}
Substituting equation $(4)$ into equation $(3)$, and expanding up to third order, we get
\begin{eqnarray}
&&\hspace*{-1.3cm}\frac{\partial^2 \phi}{\partial x^2}=-n_{cp}-\mu_1+\mu_2-\mu_3+\gamma_1 \phi+\gamma_2 \phi^2 \nonumber\\
&&\hspace*{2.4cm}+\gamma_3 \phi^3+\cdot\cdot\cdot\cdot\cdot\cdot\cdot\cdot\cdot\cdot, \label{Ch4}
\end{eqnarray}
where
\begin{eqnarray}
&&\hspace*{-3.0cm}\gamma_1=(1-\beta)(\mu_1\sigma _1+\mu_2\sigma _2), \nonumber\\
&&\hspace*{-3.0cm}\gamma_2=(\mu^2_2 \sigma^2_2-\mu^2_1 \sigma^2_1)/2, \nonumber\\
&&\hspace*{-3.0cm}\gamma_3=(1+3\beta)(\mu_1 \sigma^3_1+\mu_2 \sigma^3_2)/6,  \nonumber\
\end{eqnarray}
and
~~$\sigma_1=\frac{T_{eff}}{T_{hp}}$,~~~~~$\sigma_2=\frac{T_{eff}}{T_{e}}$,~~~~~$\mu_1=\frac{n_{hp0}}{n_{cp0}}$,

~~~~$\mu_2=\frac{n_{e0}}{n_{cp0}}$,~~~~~$\mu_3=\frac{n_{i0}}{n_{cp0}}$,~~~~~$T_{eff}=\frac{T_e T_{hp}}{\mu_1 T_e+\mu_2 T_{hp}}$.

\noindent In the above equations, the cold positron number density $n_{cp}$ is  normalized by its unperturbed number density $n_{cp0}$ ; $u_{cp}$ is the cold positron
fluid speed normalized by the PA wave speed $C_{cp}=( k_B T_{eff}/m_p)^{1/2}$; $\phi$ is
the electrostatic wave potential normalized by $k_BT_{eff}/e$; where $k_B$ being the
Boltzmann constant, $T_{eff}$ being the effective temperature,
$m_p$ being the positron rest mass, and $e$ being the magnitude of single electron
charge. The time and space variables are normalized
by ${\omega^{-1}_{cp}}=(m_p/4\pi e^2 n_{cp0})^{1/2}$ and $\lambda_{Dp}=(k_BT_{eff}/4 \pi  e^2  n_{cp0})^{1/2}$ respectively.
%%%%%%%%%%%%%%%%%%%%%%%%%%%%%%%%%%%%%%%%%%%%%%%%%%%%%%%%%%%%%%%%%%%%%%%%%%%%%%%%%%%%%%%%%%%%%
\begin{figure*}[htp]
  \centering
  \begin{tabular}{ccc}
    \includegraphics[width=80mm]{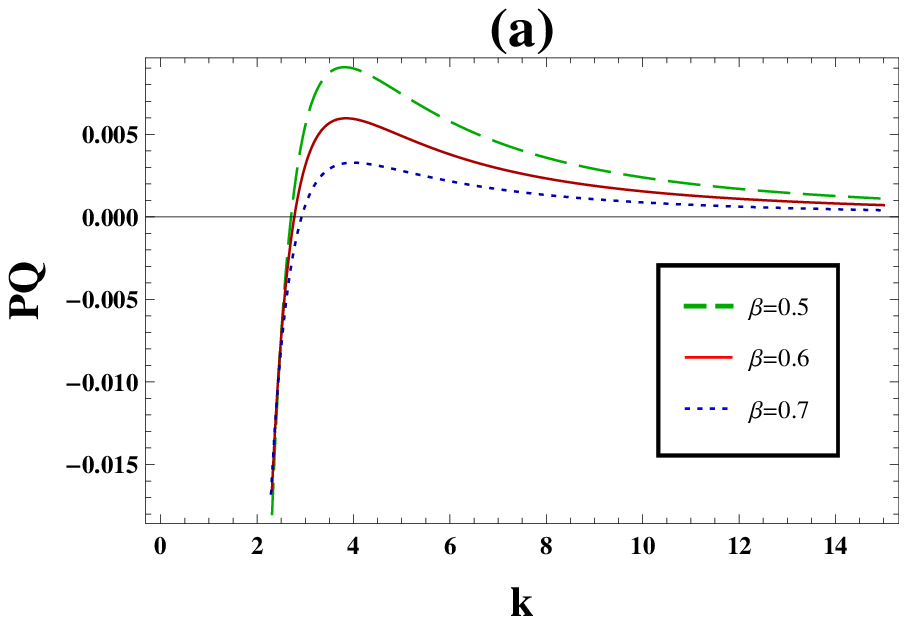}&
    \includegraphics[width=80mm]{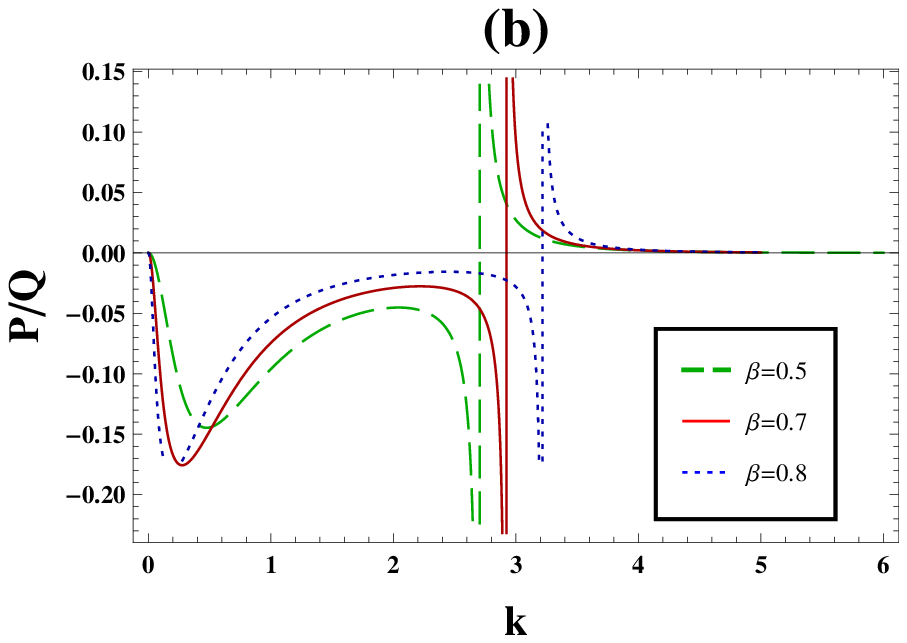}\\
  \end{tabular}
  \label{figur}\caption{(Color online)The Variation of $PQ$ or $P/Q$  with wave number $k$ for different values of $\beta$.
  (a) $PQ$  against $k$ for $\beta$, (b) $P/Q$  against $k$ for $\beta$. All the figures are generated by using these values, $\mu_1=0.2,
  \mu_2=0.7, \sigma_1=3, \sigma_2=1.5$ and $\beta=0.5$.}
\end{figure*}
\begin{figure*}[htp]
  \centering
  \begin{tabular}{ccc}
    \includegraphics[width=80mm]{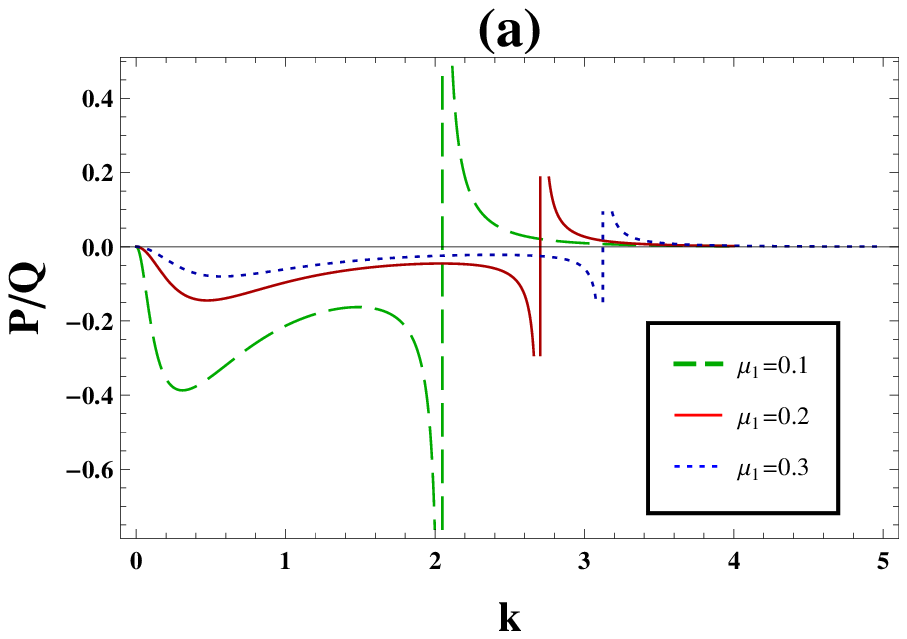}&
    \includegraphics[width=80mm]{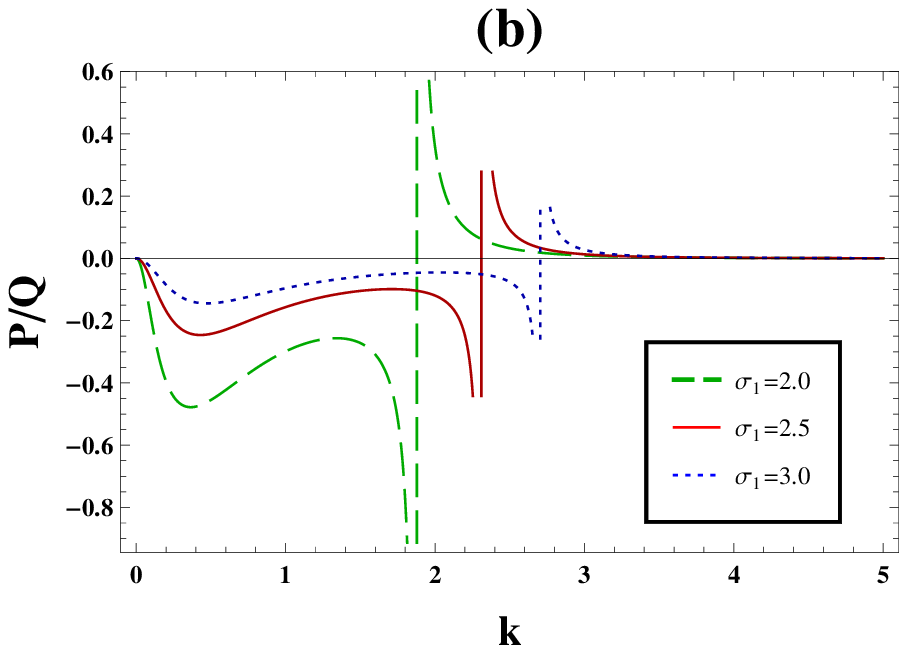}\\

    \includegraphics[width=80mm]{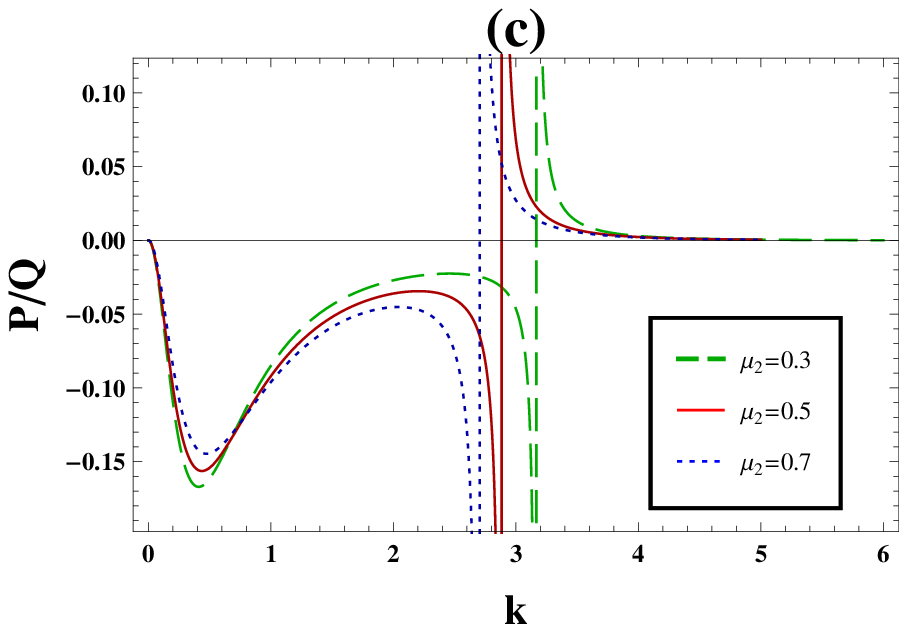}&
    \includegraphics[width=80mm]{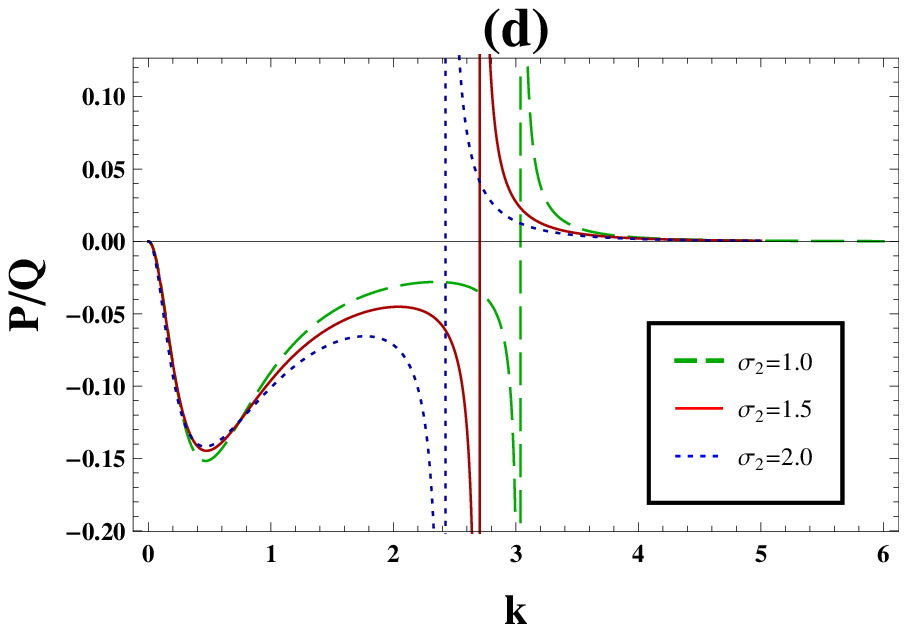}\\
  \end{tabular}
  \label{figur}\caption{(Color online) The Variation of $P/Q$ against $k$ for different values of plasma parameters.
  (a) For $\mu_1$, (b) For $\sigma_1$, (c) For $\mu_2$ and (d) For $\sigma_2$.}
\end{figure*}

\begin{figure*}[htp]
  \centering
  \begin{tabular}{ccc}
    \includegraphics[width=55mm]{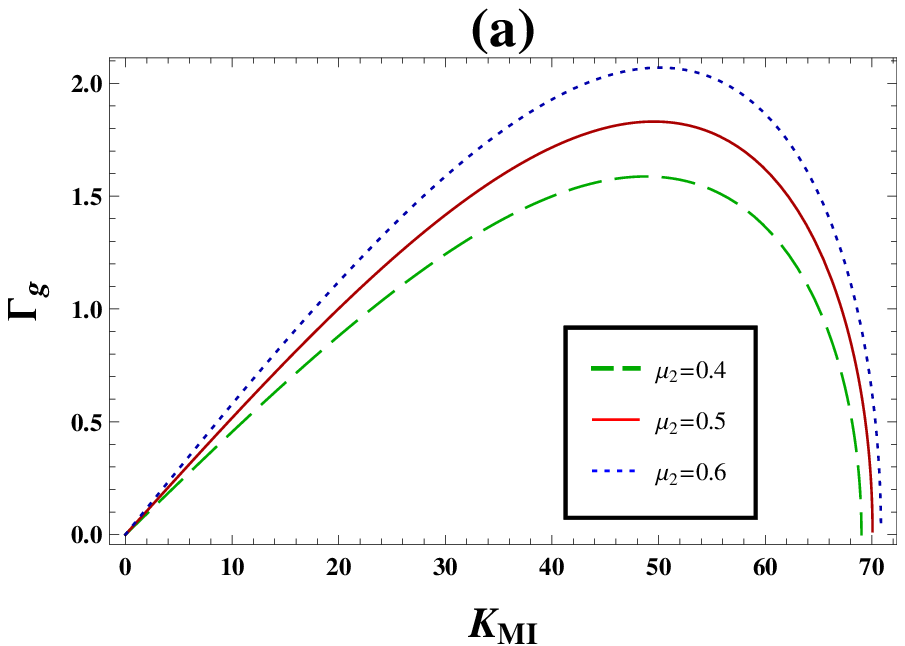}&
    \includegraphics[width=55mm]{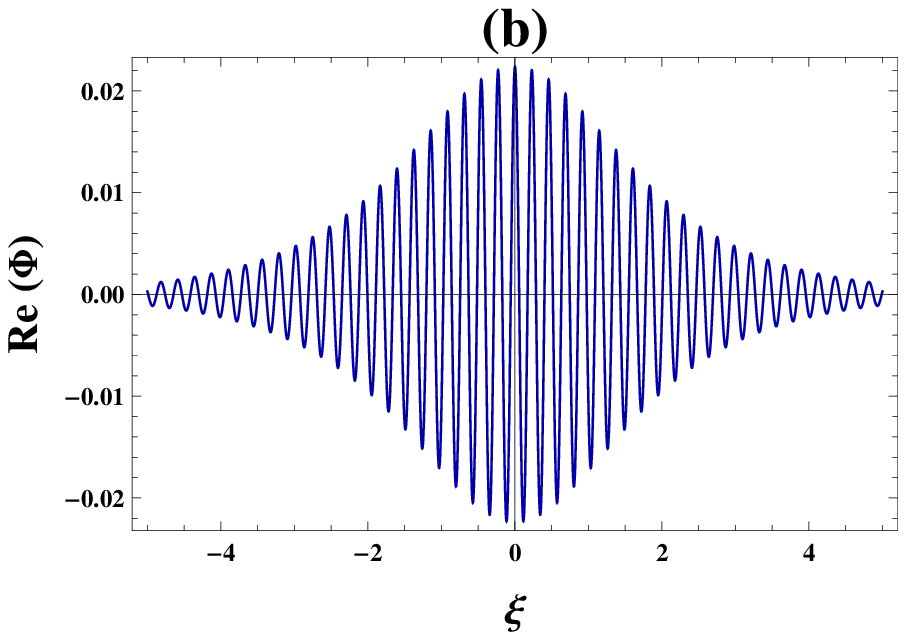}&
    \includegraphics[width=55mm]{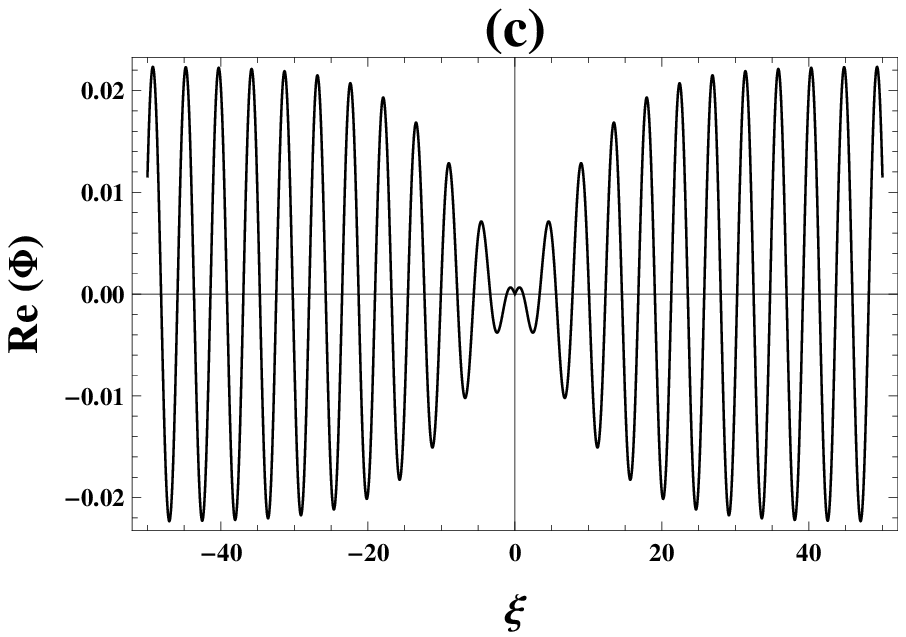}\\
  \end{tabular}
  \label{figur}\caption{(Color online) Plot of the MI growth rate $(\Gamma_g)$ against ${k_{MI}}$ for different values of $\mu_2$, along with $k=6$, and $\Phi_{0}=0.6$.
  (b) Bright envelope solitons for $k=5$, (c) Dark envelope solitons for $k=2.2$.
  Along with $\psi_{0}=0.0005, U=0.1, \tau=0,$ and  $\Omega_0=0.4$.}
\end{figure*}
%%%%%%%%%%%%%%%%%%%%%%%%%%%%%%%%%%%%%%%%%%%%%%%%%%%%%%%%%%%%%%%%%%%%%%%%%%%%%%%%%%%%%%%%
\section{Derivation of the NLS equation}
To study the modulation of the PAWs in our considered plasma system, we will derive the NLS equation
 by employing the reductive perturbation method. So we first introduce the independent variables
are stretched as
\begin{eqnarray}
&&\hspace*{-3.7cm}\xi={\epsilon}(x  - v_gt),~~~\tau={\epsilon}^2 t, \label{eq11}
\end{eqnarray}
where $v_g$ is the envelope group velocity to be determined later and $\epsilon ~(0<\epsilon<1)$ is a small
(real) parameter. Then we can write a general expression for the dependent variables as
\begin{eqnarray}
&&\hspace*{-0.5cm}M(x,t)=M_0 +\sum_{m=1}^{\infty}\epsilon^{(m)}\sum_{l=-\infty}^{\infty}M_{l}^{(m)}(\xi,\tau)~\mbox{exp}(i l\Theta), \nonumber\\
&&\hspace*{-0.5cm}  M_l^{(m)}=[n_{pcl}^{(m)}, u_{pcl}^{(m)}, \phi_l^{(m)}]^T,~~M_l^{(0)}=[1, 0, 0]^T,\label{eq11}
\end{eqnarray}
where  $\Theta=kx-\omega t$, where k and $\omega$ are real variables representing the carrier
wave number and frequency, respectively.  $ M_l^{(m)}$ satisfies the pragmatic condition $
M_l^{(m)}= M_{-l}^{(m)^*}$, where the asterisk denotes the complex conjugate. The derivative operators
 in the above equations are treated as follows:
\begin{eqnarray}
&&\hspace*{-01.0cm}\frac{\partial}{\partial t}\rightarrow\frac{\partial}{\partial t}-\epsilon v_g \frac{\partial}{\partial\xi}+\epsilon^2\frac{\partial}{\partial\tau},~~~~\frac{\partial}{\partial x}\rightarrow\frac{\partial}{\partial x}+\epsilon\frac{\partial}{\partial\xi}. \label{eq10}
\end{eqnarray}
Substituting equations $(6)-(8)$ into equations $(1),(2)$, and $(5)$ and collecting the power
terms of $\epsilon$, the first order  ($m=1$) equation with ($l=1$) give
\begin{eqnarray}
&&\hspace*{-1.0cm}-i\omega n_1^{(1)}+iku_1^{(1)}=0,~~-i\omega u_1^{(1)}+ik\phi_1^{(1)}=0,\nonumber\\
&&\hspace*{-1.0cm}n_1^{(1)}-k^2\phi_1^{(1)}-\gamma_1\phi_1^{(1)}=0.\label{eq14}
\end{eqnarray}
The solution for the first harmonics read as
\begin{eqnarray}
&&\hspace*{-3.4cm} n_1^{(1)}=\frac{k^2}{\omega^2}\phi_1^{(1)},~~~ u_1^{(1)}=\frac{k}{\omega}\phi_1^{(1)}.\label{eq11}
\end{eqnarray}
We thus obtain the dispersion relation for PAWs
\begin{eqnarray}
&&\hspace*{-5.5cm} \omega^2=\frac{k^2}{(k^2+\gamma_1)}.\label{eq12}
\end{eqnarray}
The second-order when $(m=2)$ reduced equations with $(l=1)$ are
\begin{eqnarray}
&&\hspace*{-2.3cm}n_1^{(2)}=\frac{k^2}{\omega^2}\phi_1^{(2)}+\frac{2ik(v_g k-\omega)}{\omega^3} \frac{\partial \phi_1^{(1)}}{\partial\xi},\nonumber\\
&&\hspace*{-2.3cm}u_1^{(2)}=\frac{k}{\omega}\phi_1^{(2)}+\frac{i(v_g k-\omega)}{\omega^2} \frac{\partial \phi_1^{(1)}}{\partial\xi},\label{eq16}
\end{eqnarray}
with the compatibility condition
\begin{eqnarray}
&&\hspace*{-4.3cm}v_g=\frac{\partial \omega}{\partial k}=\frac{\omega (1-\omega^2)}{k}.\label{eq17}
\end{eqnarray}
The amplitude of the second-order harmonics are found to be proportional to $|\phi_1^{(1)}|^2$
\begin{eqnarray}
&&\hspace*{-1.9cm}n_2^{(2)}=C_1|\phi_1^{(1)}|^2,~~~~~n_0^{(2)}=C_4 |\phi_1^{(1)}|^2,\nonumber\\
&&\hspace*{-1.9cm}u_2^{(2)}=C_2 |\phi_1^{(1)}|^2 ,~~~~~u_0^{(2)}=C_5 |\phi_1^{(1)}|^2,\nonumber\\
&&\hspace*{-1.9cm}\phi_2^{(2)}=C_3 |\phi_1^{(1)}|^2,~~~~~\phi_0^{(2)}=C_6|\phi_1^{(1)}|^2,\label{eq19}
\end{eqnarray}
where
\begin{eqnarray}
&&\hspace*{-3.4cm}C_1=\frac{3k^4}{2\omega^4}+\frac{C_3 k^2}{\omega^2},\nonumber\\
&&\hspace*{-3.4cm}C_2=\frac{k^3}{2\omega^3}+\frac{C_3 k}{\omega},\nonumber\\
&&\hspace*{-3.4cm}C_3=\frac{3 k^4-2\gamma_2 \omega^4}{2 \omega^4(4 k^2+\gamma_1)-2\omega^2 k^2}. \nonumber\\
&&\hspace*{-3.4cm}C_4=\frac{2v_g k^3 +\omega k^2+C_6\omega^3 }{v_g^2\omega^3 },\nonumber\\
&&\hspace*{-3.4cm} C_5=\frac{k^2+C_{6}\omega^2 }{v_g\omega^2 },\nonumber\\
&&\hspace*{-3.4cm}C_6=\frac{2v_g k^3+\omega k^2-2\gamma_2 v_g^2 \omega^3}{ \gamma_1 v_g^2\omega^3-\omega^3}.\nonumber\
\end{eqnarray}
Finally, the third harmonic modes $(m=3)$ and $(l=1)$ and  with the help of  equations $(10) - (14)$,
give a system of equations, which can be reduced to the following  NLS equation:
\begin{eqnarray}
&&\hspace*{-3.7cm}i\frac{\partial \Phi}{\partial \tau}+P\frac{\partial^2 \Phi}{\partial \xi^2}+Q|\Phi|^2\Phi=0, \label{eq24}
\end{eqnarray}
where $\Phi=\phi_1^{(1)}$ for simplicity. The dispersion coefficient $P$ is
\begin{eqnarray}
&&\hspace*{-4.5cm}P=\frac{1}{2}\frac{\partial v_g}{\partial k}=-\frac{3}{2} \frac{\omega^2}{k} v_g,\nonumber\
\end{eqnarray}
and the nonlinear coefficient Q is
\begin{eqnarray}
&&\hspace*{-1.4cm}Q=\frac{\omega^3}{2 k^2}\left[-\frac{k^2(C_1+C_4)}{\omega^2}+2\gamma_2(C_3+C_6)\right.\nonumber\\
&&\hspace*{1.5cm}\left.+3\gamma_3-\frac{2k^3(C_2+C_5)}{\omega^3}\right].\nonumber\
\end{eqnarray}
%%%%%%%%%%%%%%%%%%%%%%%%%%%%%%%%%%%%%%%%%%%%%%%%%%%%%%%%%%%%%%%%%%%%%%%%%%%%%.
\section{Stability analysis}
The evolution of PAWs are governed by the equation $(15)$ essentially depends on the coefficients
product $PQ$. Let us consider the harmonic modulated amplitude solution $\Phi=\Phi_o \exp(iQ{|\Phi_o|^2}\tau)$.
Following the standard stability analysis, one may perturb the amplitude by
setting $\Phi=\hat{\Phi}_0+\epsilon \hat{\Phi}_{1,0} \exp[i({k_{MI}}\xi-{\omega_{MI}}\tau)]+c.c$ (the perturbation
wave number ${k_{MI}}$ and the frequency ${\omega_{MI}}$). Hence, the nonlinear dispersion relation for the
 amplitude modulation \cite{Schamel2002,Sabry2008} is given by
\begin{eqnarray}
&&\hspace*{-2.2cm}{\omega^2_{MI}}=P^2{k^2_{MI}} \left ({k^2_{MI}}-2\frac{Q}{P}{|\Phi_o|^2}\right).\label{eq12}
\end{eqnarray}
Clearly, if $PQ<0$, ${\omega_{MI}}$ is always real for all values of ${k_{MI}}$, hence in this region the PAWs are
stable in the presence of small perturbation. On the other hand, when $PQ>0$ , the
MI would set in as ${\omega_{MI}}$ becomes imaginary and the PAWs are unstable for ${k_{MI}}<k_c=\sqrt{2Q{|\Phi_o|}^2/P}$,
where $k_c$ is the critical value of the wave number of modulation and $\Phi_o$ is the amplitude of the carrier waves.
The growth rate ($\Gamma_g$) of  MI (within this conditions, when  $PQ>0$ and simultaneously ${k_{MI}}<k_c$)  is given by
\begin{eqnarray}
&&\hspace*{-3.8cm}\Gamma_g=|P|~{k^2_{MI}}\sqrt{\frac{k^2_{c}}{k^2_{MI}}-1}.\label{eq112}
\end{eqnarray}
Clearly, the maximum value $\Gamma_{g(max)}$ of $\Gamma_g$ is obtained at ${k_{MI}}=k_c/\sqrt{2}$ and is given by $\Gamma_{g(max)}=|Q||\Phi_0|^2$.

The coefficients of dispersion term $P$ and nonlinear term $Q$
are dependent on  various physical plasma parameters, such as
$\sigma_1,~\sigma_2,~\mu_1,~\mu_2,$ and $\beta $. Thus, these
parameters may be sensitive to change the stability conditions of
the PAWs. One can recognize the stability conditions of PAWs by depicting
$P/Q$ against $k$ for different physical plasma parameters. The
stability of the profile is depicted in Figs. $1(b)$ and $2$, where it is shown that the variation of the
ratio of $P/Q$ versus $k$ for different plasma parameters. When
the sign of the ratio $P/Q$ is negative, the PAWs are modulationally
stable, while the sign of the ratio $P/Q$ is positive,
the PAWs will be  modulationally unstable against external
perturbations. It is clear that both stable and unstable region for PAWs
are obtained from the Figs. $1(s)$ and $2$. When $P/Q\rightarrow\pm\infty$,
the corresponding value of $k(=k_c)$ is called critical or
threshold wave number for the onset of MI. This critical value
separates the  unstable ($P/Q>0$) from the stable region
($P/Q<0$) one.

The nonthermal parameter plays a significant role to change the
stability of the PAWs. With the increasing values of $\beta$
(nonthermality), the critical value $k_c$ is shifted to the
higher value (see Fig. $1(b)$). It is seen that the instability sets
increases with the increasing values of nonthermal parameter
$\beta$. It is also found that the absolute value of the ratio
$P/Q$ increases with the increasing values of $\beta$.

Figure $2(a)$ shows the variation of $P/Q$ with $k$ for different
values of hot to cold positron concentration ratio (via $\mu_1$)
with fixed values of other physical parameters. The critical
value of wave number at which the instability sets increases with
the increasing values of $\mu_1$. Actually, increasing the values
of hot to cold positron concentration ratio is responsible for the
decreasing values of the nonlinear coefficient $Q$. As the reason
of the values of decreasing nonlinear coefficient the instability
sets increases. On the other hand, the absolute value of the
ratio $P/Q$ increases with the increasing values of $\mu_1$.

We have also analyzed the effect of effective temperature to the
hot positron temperature ratio (via $\sigma_1$) on the stability
of the wave profiles [see Fig. $2(b)$]. It is observed that the
decreasing values of the hot positron temperature (which leads the
increasing values of $\sigma_1$) the instability sets increases.
On the other hand, with the decreasing values of the hot electron
temperature (which leads the increasing values of $\sigma_2$) the
instability sets decreases [see Fig.$2(d)$]. So increasing hot positron or
hot electron temperature plays simultaneously opposite role to
recognize the stability region for the PAWs  to increase or decrease.

The variation of $P/Q$ with $k$ for different values of hot
electron to cold positron concentration ratio (via $\mu_2$) with
fixed values of other physical parameters is depicted in Fig. $2(c)$.
It is seen that the critical wavenumber decreases with the
increasing values of $\mu_2$. This leads that the increasing of
electron concentration, the critical value ($k_c$) is shifted to
the lower value. So excess number of electron of the system is
caused to minimize the stability of the wave profile. Again, the
absolute value of the ratio $P/Q$ decreases with the increasing
values of $\mu_2$.

The variation of MI growth rate (via $\Gamma_g$) versus MI wave
number (via $k_{MI}$) is depicted in Fig. $3(a)$. It is observed that
the growth rate increases with the increasing values of hot
electron to cold positron concentration ratio (via $\mu_2$). This
outcome also implies that the greater (lower) the values of hot electron (cold positron)
concentration, the nonlinearity of the PAWs is stimulated (depressed) which expose via the maximum value of MI
growth rate. So $\mu_2$ plays a vital role to the stability of PAWs profile.
%%%%%%%%%%%%%%%%%%%%%%%%%%%%%%%%%%%%%%%%%%%%%%%%%%%%%%%%%%%%%%%%%%%%%%%%%%%%%.
\section{Envelope solitons}
If $PQ<0$, the modulated envelope pulse is stable (in this region, dark envelope solitons exist) and
when $PQ>0$, the modulated envelope pulse is unstable against
external perturbations and leads to generation of  bright envelope
solitons. The variation of $PQ$ versus k is depicted in Fig. $1(a)$
with different values of nonthermal parameter $\beta$. It is found
that with the increasing values of nonthermal parameter $\beta$,
the values of critical wave number (when $PQ=0)$ increases. A solution of
$(15)$ may be sought in the form $\Phi=\sqrt{\psi}~\mbox{exp}(i\theta)$,
where $\psi$ and $\theta$ are real variables which are determined by substituting into the NLS
equation and separating real and imaginary parts. An interested reader is referred to
\cite{Fedele2002,Sukla2002,Kourakis2005,Shalini2015,Omar2015} for details. The different types of solution
thus obtained are clearly summarized in the following paragraphs.
\subsection{Bright solitons}
When $PQ>0$, we find bright envelope solitons. The general analytical form of bright solitons reads
\begin{eqnarray}
&&\hspace*{-10mm}\psi=\psi_{0}~ \mbox{sech}^2 \left(\frac{\xi-U\tau}{W}\right),\nonumber\\
&&\hspace*{-10mm}\theta=\frac{1}{2P} \left[U \xi+ \left(\Omega_0-\frac{U^2}{2}\right)\tau\right].\label{eq112}
\end{eqnarray}
Here, $U$ is the propagation speed (a constant), $W$ is the soliton width, and $\Omega_0$ oscillating
frequency for $U=0$. Fig. $3(b)$ represents the bright envelope solitons.
\subsection{Dark solitons}
When $PQ<0$, we find dark envelope solitons  whose general analytical form  reads as
\begin{eqnarray}
&&\hspace*{-10mm}\psi=\psi_{0}~ \mbox{tanh}^2 \left(\frac{\xi-U\tau}{W}\right),\nonumber\\
&&\hspace*{-10mm}\theta=\frac{1}{2P}\left[ U \xi-\left( \frac{U^2}{2}-2PQ\psi_{0}\right)\tau \right].
\end{eqnarray}
Interestingly, in both of the latter two equations, the relation between soliton width $W$ and the constant maximum
amplitude $\psi_{0}$ are related by
\begin{eqnarray}
&&\hspace*{-4.2cm}W=\sqrt{\frac{2|P/Q|}{ \psi_{0}}}.
\end{eqnarray}
Figure $3(c)$ represents the dark envelope solitons.
%%%%%%%%%%%%%%%%%%%%%%%%%%%%%%%%%%%%%%%%%%%%%%%%%%%%%%%%%%%%%%%%%%%%%%%%%%%%%%%%%%%%%
\section{Conclusion}
We considered an unmagnetized four component e-p-i plasma system
consisting of immobile positive ions, inertial mobile cold
positrons, and nonthermally distributed hot positrons and hot
electrons. The well-known reductive perturbation method has been used to drive a NLS
equation, which is valid for a small but finite  amplitude limit.
We have observed the existence of both stable and unstable
regions of PAWs and  how the related physical plasma parameters (hot positrons
temperature, hot electron temperature, cold electron number density,
positron number density and nonthermallity) influence to change  the stability conditions of PAWs, MI
growth rate, and formation of envelope solitons. Like with the increasing of hot positron (hot electron)
concentration, the critical value is shifted to the higher (lower) values of $k$. It is needed to highlights
here that the findings of our present investigation should be
useful for understanding the striking features of space environments (like cluster explosions, active
galactic nuclei, auroral acceleration regions, lower part of
magnetosphere, ionosphere etc.) and laboratory plasmas.

\section*{Acknowledgment}
N. A. Chowdhury is grateful to the Bangladesh Ministry of Science
and Technology for awarding the National Science and Technology
(NST) Fellowship.

\ifCLASSOPTIONcaptionsoff
  \newpage
\fi


\begin{thebibliography}{1}

\bibitem{Begelman1984} M. C. Begelman, R. D. Blanford, and M. J. Rees, Rev. Mod. Phys. {\bf 56}, 255 (1992).

\bibitem{Miller1987} H. R. Miller and P. J. Wiita, \textit{Active galactic nuclei} (Springer, Berlin, 1987).

\bibitem{Tribeche2009} M. Tribeche, K. Aoutou, S. Younsi, and R. Amour, Phys. Plasmas {\bf 16}, 072103 (2009).

\bibitem{Michel1991} F. C. Michel, \textit{Theory of Neutron Star Magnetosphere} (Chicago University Press, Chicago, 1991).

\bibitem{Hossen2014a} M. R. Hossen and A. A. Mamun, Braz. J. Phys. {\bf 44}, 673 (2014).

\bibitem{Hossen2014b} M. R. Hossen, S. A. Ema, and A. A. Mamun, Commun. Theor. Phys. {\bf 62}, 888 (2014).

\bibitem{Misner1973} W. Misner, K. S. Thorne, and J. I. Wheeler, \textit{Gravitation} (Freeman, San Francisco, 1973).

\bibitem{Rees1971} M. J. Rees, Nature {\bf 229}, 312 (1971).

\bibitem{Liang1998} E. P. Liang, S. C. Wilks, and M. Tabak, Phys. Rev. Lett. {\bf 81}, 4887 (1998).

\bibitem{Michel1982} F. C. Michel, Rev. Mod. Phys. {\bf 54}, 1 (1982).

\bibitem{Burns1983} M. L. Burns, A. K. Harding, and R. Ramaty, \textit{Positron-electron pairs in astrophysics} (American Institute of Physics, Melville, New York, 1983).

\bibitem{Goldreich1969} P. Goldreich and W. H. Julian, Astrophys. J. {\bf 157}, 869 (1969).

\bibitem{Hansen1988} E. Tandberg-Hansen and A. G. Emslie, \textit{The Physics of Solar Flares} (Cambridge University Press, Cambridge, 1988).

\bibitem{Bains2010} A. S. Bains, A. P. Misra, N. S. Saini, and T. S. Gill, Phys. Plasmas {\bf 17}, 01203 (2010).

\bibitem{Rehman2016} M. A. Rehnam, and M. P. Misra, Phys. Plasmas {\bf 23}, 012302 (2016).

\bibitem{Lu2015} L. Tie-Lu, W. Yun-Liang, and L. Yan-Zhen, Chin. Phys. B {\bf 24}, 025202 (2015).

\bibitem{Sultana2011} S. Sultana, and I. Kourakis, Plasma Phys.Control. Fusion {\bf 53}, 045003 (2011).

\bibitem{Demiray2016} H. Demiray, Phys. Plasmas {\bf 23}, 032109 (2016).

\bibitem{Haque2003} Q. Haque, and H. Saleem, Phys. Plasmas {\bf 10}, 3793 (2003).

\bibitem{Shah2010} A. Shah, R. Saeed, and M. Noaman-Ul-Haq, Phys. Plasmas {\bf 17}, 072307 (2010).

\bibitem{Sabry2009} R. Sabry, W. M. Moslem, and P. K. Sukla, Eur. Phy. J. D {\bf 51}, 233 (2009).

\bibitem{Akbari2010} M. Akbari-Moghshanjoughi, Phys. Plasmas {\bf 17}, 082315 (2010).

\bibitem{El-Awady2010} E. I. El-Awady, S. A. El-Tantawy, W. M. Moslem, and P. K. Sukla, Phys. Lett. A {\bf 374}, 3216 (2010).

\bibitem{Alam2013} M. S. Alam, M. M. Masud, and A. A. Mamun, Chin. Phys. B {\bf 22}, 115202 (2013).

\bibitem{Matsumoto1994} H. Matsumoto, H. Kojima, T. Miyatake, I. Nagano, A. Fujita, L. A. Frank, T. Mukai, W. R. Paterson, Y. Saito, S. Machida, and
                              R. R. Anderson, Geophys. Res. Lett. {\bf 21}, 2915 (1994).

\bibitem{Bostrom1992} R. Bostr\"{o}m, IEEE Trans. Plasma Sci. {\bf 20}, 756 (1992).

\bibitem{Lundin1989} R. Lundin, A. Zakharov, R. Pellinin, H. Borg, B. Hultqvist, N. Pissarenko, E. M. Dubinin, S. W. Barabash, I. Liede, and H.
                                 Koskinen, Nature  {\bf 341}, 609 (1989).

\bibitem{Dovner1994} P. O. Dovner, A. I. Eriksson, R. Bostr\"{o}m, and B. Holback, Geophys. Res. Lett. {\bf 21}, 1827 (1994).

\bibitem{Cairns1995} R. A. Cairns, A. A. Mamun, R. Bingham, R. Bostr\"{o}m, R. O. Dendy, C. M. C. Nairn, and P. K. Shukla, Geophys. Res. Lett. {\bf 22}, 2709 (1995).

\bibitem{Chatterjee2012} P. Chatterjee, D. K. Ghosh, and B. Sahu, Astrophys. Space Sci. {\bf 339}, 261 (2012).

\bibitem{Pakzad2009} H. R. PAkzad, Phys. Lett. A {\bf 373}, 847-850 (2009).

\bibitem{Biswajit2010} B. Sahu, Phys. Scr. {\bf 82},065504 (2010).

\bibitem{Messekher2016} A. Messekher, K. Mebrouk, L. A. Gougam, and M. Tribeche, Phys. Plasmas  {\bf 23}, 104504 (2016).

\bibitem{Eslami2011} P. Eslami, M. Mottaghizadeh, and H. R. Pakzad, Phys. Plasmas  {\bf 18}, 102313 (2011).

\bibitem{Zhang2009} J. Zhang, Y. Wang, and L. Wu, Phys. Plasmas  {\bf 16}, 062102 (2009).

\bibitem{Schamel2002} R. Fedele and H. Schamel, Eur. Phys. J. B  {\bf27}, 313 (2002).

\bibitem{Sabry2008} R. Sabry,  Phys. Plasmas  {\bf15}, 092101 (2008).

\bibitem{Fedele2002} R. Fedele, Phys. Scr. {\bf65}, 502 (2002).

\bibitem{Sukla2002} R. Fedele, H. Schamel, and P.K. Sukla, Phys. Scr.  {\bf T98}, 18 (2002).

\bibitem{Kourakis2005} I. Kourakis and P.K. Sukla, Nonlinear Proc. Geophys.  {\bf 12}, 407 (2005).

\bibitem{Shalini2015} Shalini, N. S. Saini, and A. P. Misra, Phys. Plasmas {\bf22}, 092124 (2015).

\bibitem{Omar2015} O. Bouzit, M. Tribeche, and A. S. Bains, Phys. Plasmas {\bf22}, 084506 (2015).

\end{thebibliography}
\end{document}